\documentstyle[12pt]{article}

\def\L{{\cal L}}
\def\M{{\cal M}}

\def\Res{\mathop{\rm Res}\limits}

\def\endproof{\hbox to \hsize{\hfil $\Box$}}
\def\endexample{\centerline{\vbox{\hrule width4in}}}

\begin{document}
\hbox to \hsize{\hfil DTP 96/13}
\hbox to \hsize{\hfil May, 1966}
\bigskip

\bigskip
\centerline{{\Large\bf The algebraic and Hamiltonian structure of the}}

\vspace{.2in}
\centerline{{\Large\bf dispersionless Benney and Toda hierarchies}}

\vspace{.4in}
\centerline{{\bf D.B. Fairlie}}

\vspace{.2in}
\centerline{Dept. of Mathematical Sciences, University of Durham,}
\vspace{.1in}
\centerline{Durham, DH1 3LE, England
\footnote{e-mail: david.fairlie@durham.ac.uk}.}

\vspace{.3in}
\centerline{{\bf I.A.B. Strachan}}\vspace{.1in}
\centerline{Dept. of Pure Mathematics and Statistics, University of Hull,}
\vspace{.1in}
\centerline{Hull, HU6 7RX, England
\footnote{e-mail: i.a.b.strachan@hull.ac.uk}.}

\vspace{.4in}
\centerline{{\bf Abstract}}

\vspace{.3in}
\small
\parbox{5.8in}{The algebraic and Hamiltonian structures of the
multicomponent dispersionless Benney and Toda hierarchies are studied.
This is achieved by using a modified set of variables for which there
is a symmetry between the basic fields. This symmetry enables formulae
normally given implicitly in terms of residues, such as conserved charges
and fluxes, to be calculated explicitly. As a corollary of these results
the equivalence of the Benney and Toda hierarchies is established. It is
further shown that such quantities may be expressed in terms of
generalized hypergeometric functions, the simplest example involving
Legendre polynomials. These results are then extended to systems derived
from a rational Lax function and a logarithmic function.
Various reductions are also studied.}
\normalsize

\bigskip

\section*{1. Introduction }

\bigskip

One of the most studied integrable systems is the Toda lattice equation

\[
\frac{\partial^2 \rho_n}{\partial t \partial {\tilde t} } =
e^{\rho_{n+1}} - 2 e^{\rho_n} + e^{\rho_{n-1}}\,,
\]

\noindent together with its various generalisations. Besides the inherent
interest in such integrable systems, these Toda systems have a
wide number of applications, form sorting theory to quantum field
theory and differential
geometry. The one dimensional Toda chain 

\begin{equation}
\frac{\partial^2\rho_n}{\partial t^2} =
e^{\rho_{n+1}} - 2 e^{\rho_n} + e^{\rho_{n-1}}\,,
\label{eq:2dtodalattice}
\end{equation}

\noindent may be written as a coupled system of difference equations

\begin{equation}
\begin{array}{rcl}
S_{n,t} & = & (\Delta-1) P_n \,, \\
P_{n,t} & = & P ( 1 - \Delta^{-1}) S_n \,,
\end{array}
\label{eq:discretetoda}
\end{equation}

\noindent where $\rho_n=\log P_n$ and $\Delta$ is the shift operator
defined by $\Delta^\pm \rho_n = \rho_{n\pm 1}\,,$ and the properties
of such systems have been extensively studied, notably by Kuperschmidt
\cite{Kuperschmidt}.

\bigskip

This paper concerns a particular limit of these systems, the continuum or
long wave limit. The idea is to look for solutions whose natural length
scale is large compared with the lattice spacing, or, alternatively, one
lets the lattice spacing tend to zero. In the limit the discrete label
$n$ becomes a continuous variable $x$ and the operator
$\Delta^{+1}-2+\Delta^{-1}$
degenerates into a second derivative. Thus under this limit the
two dimensional
Toda lattice (\ref{eq:2dtodalattice}) becomes

\begin{equation}
\frac{\partial^2 \rho}{\partial t \partial {\tilde t}} =
\frac{\partial^2\, e^\rho}{\partial x^2}\,.
\label{eq:BoyerFinley}
\end{equation}

\noindent One intriguing property of these systems is that the notion of
integrability is preserved in this limit, so (\ref{eq:BoyerFinley}) is in
fact a multidimensional integrable system. In this paper this
limit will be called the dispersionless limit, as
mathematically it is the same as the
limit which send the KdV equation into the Monge equation

\begin{eqnarray*}
u_t & = & u u_x + \kappa u_{xxx} \,, \\
& \downarrow & \kappa \rightarrow 0  \\
u_t & = & u u_x\,.
\end{eqnarray*}

\noindent Many of the geometrical properties of these integrable
systems have been derived in \cite{TakasakiTakebe}.
One of the more recent motivations for the
study of such integrable systems has come from their role in topological
field theory \cite{Dubrovin}.

\bigskip

Under this limit the difference system (\ref{eq:discretetoda}) becomes

\begin{equation}
\begin{array}{rcl}
S_t & = & P_x \,, \\
P_t & = & P S_x \,.
\end{array}
\label{eq:toda}
\end{equation}

\noindent and in an earlier paper \cite{FS} this system
was studied. This was achieved by transforming to a set of
modified variables defined by

\begin{eqnarray*}
S & = & u+v \,, \\
P & = & uv\,.
\end{eqnarray*}

\noindent In these new variables the Toda system (\ref{eq:toda}) becomes
symmetric:

\begin{equation}
\begin{array}{rcl}
u_t & = & u v_x \,, \\
v_t & = & v u_x
\end{array}
\label{eq:mtoda}
\end{equation}

\noindent and various properties of this system, such as its associated
hierarchy and Hamiltonian structure, were studied. In this paper
these result are extended to the multicomponent Toda hierarchy.

\bigskip

The multicomponent Toda hierarchy is defined in terms of a Lax function

\[
\L(p) = p^{N-1} + \sum_{i=-1}^{N-2} p^i S_i(x,{\bf t})\,,
\quad\quad{\bf t} =\{t_1\,,t_2\,,\dots \}
\]

\noindent by the Lax equation

\begin{equation}
\frac{\partial\L}{\partial t_n} = \{ (\L^{\frac{n}{N-1}})_{+} , \L \}\,.
\label{eq:lax}
\end{equation}

\noindent Here the bracket is
defined by the formula

\[
{ \{ f,g \} } = p \frac{\partial f}{\partial p}
\frac{\partial g}{\partial x} - p \frac{\partial f}{\partial x}
\frac{\partial g}{\partial p}\,.
\]

\noindent and, as usual, $( {\cal O} )_{+}$ denotes the projection of the
function $\cal O$ onto non-negative powers of $p\,.$
More general Lax functions
will be considered in section 6 and 7.

\bigskip

\noindent {\bf Example 1.1} One obtains from the
Lax equation (\ref{eq:lax})
with $N=2\,,n=1$ equation (\ref{eq:toda})
and, with $N=3\,,n=1$ the system (where $\L=p^2 + Sp + P + Q p^{-1}\,$)

\begin{equation}
\begin{array}{rcl}
S_t & = & P_x - \frac{1}{2} S S_x \,, \\
P_t & = & Q_x \,, \\
Q_t & = & \frac{1}{2} Q S_x\,.
\end{array}
\label{eq:3toda}
\end{equation}

\bigskip

\endexample

\bigskip

The modified variables alluded to above are defined by the
factorization of the Lax function

\begin{equation}
\L=\frac{1}{p} \prod_{i=1}^N [p+u_i]
\label{eq:mlax}
\end{equation}

\noindent This transformation may be thought of as a dispersionless
Miura map, and their use in the study of the dispersionless Toda
hierarchy is due to Kuperschmidt \cite{Kuperschmidt}. 
Recall that the transformation from the KdV to the modified
KdV equation comes from a similar factorization of the Lax operator for
the KdV equation. These new  variables $u_i$ will be called the modified
variables.

\bigskip

There are a number of advantages in moving to these  modified variables.
Firstly, it puts all the fields on an egalitarian footing and any formula
(such as those for the conserved charges and fluxes and Hamiltonian
structures) will be symmetric functions of these
variables (explicitly, such formulae are invariant under the map
$u_i\mapsto u_{\sigma(i)}$ where $\sigma$ is an element of the permutation
group). Secondly, it enables such functions to be calculated explicitly,
using simple combinatorial arguments. In fact, rather than the calculations
being of the form

\[
(1+{\rm sum~of~N~terms})^\alpha
\]

\noindent the calculations are of the form

\[
\sum_N (1+{\rm single~term})^\alpha
\]

\noindent and the symmetry between the modified variables means that only
a single calculation has to be performed. The {\sl existence} of such
quantities and some of their properties follows from the general theory
of these Lax equations (see, for example \cite{AoyamaKodama}, and the
references therein). However, the {\sl explicit} forms for these
quantities are rarely given except, perhaps, for the lowest members
of the hierarchy. The modified variables enables one to perform the
general calculations with an arbitrary numbers of fields.

\bigskip

One further property of these modified variables is best
illustrated by means of an example.

\bigskip

\noindent{\bf Example 1.2} In these modified variables equation
(\ref{eq:3toda}) becomes

\begin{equation}
\begin{array}{rcl}
u_t & = & \displaystyle{\frac{1}{2} u ( -u_x+v_x+w_x ) }\,,\\
v_t & = & \displaystyle{\frac{1}{2} v ( +u_x-v_x+w_x ) } \,,\\
w_t & = & \displaystyle{\frac{1}{2} w ( +u_x+v_x-w_x ) }\,.
\end{array}
\label{eq:3example}
\end{equation}

\noindent where $S=u+v+w\,,P=uv+vw+wu$ and $Q=uvw\,.$

\bigskip

It is clear from these equations that one possible reduction
of this system comes from the constraint $v=w\,.$
In the original variables this constraint is

\[
4P^3+27 Q^2 - 18 PQS - P^2 S^2 + 4 Q S^3 = 0 \,,
\]

\noindent this being the condition for the cubic equation
$\L(p)=0$ to have a double root.
In the extreme case where all the fields are equal one obtains, after
some rescalings, the dispersionless KdV hierarchy

\[
u_{t_n} = u^n u_x\,.
\]

\noindent Thus reductions are far easier to study using these modified
variables.

\bigskip

\endexample

\bigskip

\noindent Such reductions are studied in section 5.

\bigskip

This Toda hierarchy is related to  the Benney moment
equations \cite{benney} which take the form

\[
\frac{\partial {\hat S}_n}{\partial t}=
\frac{\partial {\hat S}_{n+1}}{\partial x}+
n {\hat S}_n\frac{\partial {\hat S}_0}{\partial x}\,,
\quad\quad n = 0 \,,1\,, \ldots\,.
\]

\noindent The corresponding Lax function for a truncation of
these equations at
finite $N$ is given by
\[
\L_B=p+\sum _{n=0}^{N-1}\frac{{\hat S}
_n(x,{\bf t})}{p^{n}}\,.
\]

\noindent and the corresponding Lax equation is

\[
\L_{B,t} = \{ (\L_B)_{+} , \L_B \}\,.
\]

\bigskip

\noindent{\bf Example 1.3} For $N=2$ the Benney system coincides with the
dispersionless Toda equation (\ref{eq:toda}). For $N=3$ the
Benney equations are
(where $\L_B=p+{\hat S}+{\hat P} p^{-1} + {\hat Q} p^{-2}\,$)

\begin{equation}
\begin{array}{rcl}
{\hat S}_t & = & {\hat P}_x  \,, \\
{\hat P}_t & = & {\hat P}{\hat S}_x+{\hat Q}_x \,, \\
{\hat Q}_t & = & {2} {\hat Q} {\hat S}_x\,.
\end{array}
\label{eq:benney}
\end{equation}

\bigskip

These equations possess an infinite number of conservation laws of
the form

\[
{\hat Q}^{(n)}_t={\hat F}^{(n)}_x
\]

\noindent Using the equations of motion one finds that

\begin{eqnarray*}
{\hat Q}^{(n)}_{\hat S} & = & {\hat F}^{(n)}_{\hat P} \,,\\
{\hat P} {\hat Q}^{(n)}_{\hat P} + 2 {\hat Q} {\hat Q}^{(n)}_{\hat Q}
& = &
{\hat F}^{(n)}_{\hat S} \,, \\
{\hat Q}^{(n)}_{\hat P} & = & {\hat F}^{(n)}_{\hat Q}\,,
\end{eqnarray*}

\noindent and so the charge and flux may be written in terms of a
potential ${\hat H}^{(n)}$ via

\[
Q^{(n)}=\frac{\partial H^{(n)}}{\partial {\hat  P}};\quad P^{(n)}=
\frac{\partial H^{(n)}}{\partial {\hat  S}},
\]

\noindent the first few being

\begin{eqnarray*}
{\hat H}^{(0)} & = & {\hat P}\,,\\
{\hat H}^{(1)} & = & 2 {\hat Q}+2 {\hat P}{\hat Q}\,,\\
{\hat H}^{(2)} & = & 3 {\hat P}^2 +6 {\hat Q}{\hat S} \,,\\
{\hat H}^{(3)} & = & 12 {\hat P}{\hat Q} + 12 {\hat P}^2{\hat S}+12
{\hat Q}{\hat S}^2+4{\hat P}{\hat S}^3\,.
\end{eqnarray*}

\noindent With these one obtains equations satisfied by this
potential, namely

\begin{eqnarray*}
\frac{\partial^2 {\hat H}^{(n)}}{\partial {\hat  P}^2}&=&
\frac{\partial^2 {\hat H}^{(n)}}{\partial {\hat  S}\partial {\hat  Q}}
\,,\\
\frac{\partial^2 {\hat H}^{(n)}}{\partial {\hat  S}^2}&=&
{\hat  P}\frac{\partial^2 {\hat H}^{(n)}}{\partial {\hat  P}^2}+2
{\hat  Q}
\frac{\partial^2 {\hat H}^{(n)}}{\partial {\hat  P}\partial {\hat  Q}}\,.
\end{eqnarray*}

\noindent It is also possible to derive many other equations satisfied
by this
potential, for example,

\[
{\hat  Q}\frac{\partial^2 {\hat H}^{(n)}}{\partial {\hat  Q}^2} =
\frac{1}{2}
\frac{\partial^2 {\hat H}^{(n)}}{\partial {\hat  S}\partial {\hat  P}}-
\frac{1}{2}
\frac{\partial^2({\hat P}{\hat H}^{(n)})}
{\partial{\hat P}\partial{\hat Q}}\,.
\]

\noindent Notice that if $H^{(n)}$ is a solution of the above
equations, so is

\[
{H^{(n-1)}=\frac{1}{n+1}\,\frac{\partial H^{(n)}}{\partial {\hat  S}}}\,.
\]

\noindent This implies that all lower conserved densities may be
obtained from the higher ones by
differentiation with respect to $\hat S$.
A similar observation for the KdV
conserved quantities has been noted by Gerald Watts \cite{pc}.
There is an extremely simple formula for polynomial solutions for the
potential $H^{(n)}$.
It is simply given by the coefficient of $p^{-1}$ in the expansion of

\[
(p + {\hat  S}  + \frac{\hat  P}{p} + \frac{\hat  Q}{p^2})^{n+1}
\]

\noindent i.e. as the integral

\[
{\hat H}^{(n)} = \frac{1}{2\pi i} \oint \L_B^{n+1} dp\,.
\]

\noindent These results generalize to arbitrary $N$.

\bigskip

In terms of the modified variables
(where ${\hat S} ={\hat u}+{\hat v}+{\hat w}$
etc.) this becomes

\begin{equation}
\begin{array}{rcl}
{\hat u}_t & = & {\hat u} ( {\hat v}_x+{\hat w}_x ) \,,\\
{\hat v}_t & = & {\hat v} ( {\hat w}_x+{\hat u}_x ) \,,\\
{\hat w}_t & = & {\hat w} ( {\hat u}_x+{\hat v}_x ) \,,
\end{array}
\label{eq:3benney}
\end{equation}

\noindent and the equivalence between the Toda and Benney systems is
given by the
change of variables

\[
{\hat u} = v w\,,\quad\quad{\hat v}= w u\,,\quad\quad{\hat w} = u v\,.
\]

\noindent With such a change the Benney system (\ref{eq:3benney})
transforms into

\begin{eqnarray*}
u_t & = & u (vw)_x \,, \\
v_t & = & v (wu)_x \,, \\
w_t & = & w (uv)_x \,,
\end{eqnarray*}

\noindent which is the $N=3\,,n=2$ flow of the Toda system (\ref{eq:lax}).
Without of use of modified variables such an equivalence is less clear.

\bigskip

\endexample

\bigskip

\noindent Such changes of variable are discussed in greater
depth in section 5.
This will show the equivalence between the Toda and Benney hierarchies.
An alternative way of showing this equivalence is given in section 6.

\bigskip

Throughout this paper the binomial coefficients $\pmatrix{a \cr b}$
should be
interpreted in terms of $\Gamma$-functions, i.e.

\[
\pmatrix{a \cr b} = \frac{  \Gamma(a+1) }{ \Gamma(a-b+1) \Gamma(b+1) }\,.
\]

\noindent This enables the binomial coefficients to be defined for
fractional
values of $a$ and $b\,.$ Such coefficients may also be defined
for negative
values if one manipulates the formula formally, using the basic definition
$\Gamma(z+1) = z \Gamma(z)\,.$ For example
\begin{eqnarray*}
\pmatrix{-2 \cr +2} & = & \frac{ \Gamma(-1) }{\Gamma(-3) \Gamma(3)} \,, \\
& = & \frac{(-2) \Gamma(-2) }{\Gamma(-3)\Gamma(3)} \,, \\
& = & \frac{(-2)(-3) \Gamma(-3)}{(+2)(+1)\Gamma(-3)} = 3\,.
\end{eqnarray*}
\noindent For notational simplicity binomial coefficients have
been used, rather
than the more correct $\Gamma$-function notation.

\bigskip

\section*{2. The multicomponent Toda hierarchy}

\bigskip

It follows from the general theory behind such Lax equations
\cite{Kuperschmidt,AoyamaKodama} that the
quantities

\begin{equation}
Q^{(n)} = \frac{1}{2\pi i} \oint \L^{ \frac{n}{N-1} } \frac{dp}{p}
\label{eq:charges}
\end{equation}

\noindent are conserved with respect to the evolutions defined by the
Lax equation (\ref{eq:lax}).

\bigskip

\noindent {\bf Proposition 2.1}

\bigskip

The conserved charges defined by (\ref{eq:charges}) are given by the
formula

\begin{equation}
Q^{(n)} = \sum_{ \{ r_i \,: \sum_{i=1}^N r_i= n \} }
\Bigg\{  \prod_{i=1}^N
\pmatrix{ \frac{n}{N-1} \cr r_i } u_i^{r_i} \Bigg\}\,.
\label{eq:charges2}
\end{equation}

\bigskip

\noindent{\bf Proof}

\bigskip

It follows from (\ref{eq:mlax}) that

\[
{\L}^{ \frac{n}{N-1} } = p^n \prod_{i=1}^N
\Big( 1 + \frac{u_i}{p} \Big)^{ \frac{n}{N-1} }
\]

\noindent and so, using the binomial expansion,

\[
{\L}^{ \frac{n}{N-1} } = p^n \prod_{i=1}^N
\sum_{r_i=0}^\infty
\pmatrix{ \frac{n}{N-1} \cr r_i } u_i^{r_i} p^{-r_i}\,.
\]

\noindent Using the residue theorem the
integral formula (\ref{eq:charges})
may be easily evaluated,

\begin{eqnarray*}
Q^{(n)} & = & {\rm coefficient~of~}p^{-n}{\rm~in}
\prod_{i=1}^N
\sum_{r_i=0}^\infty
\pmatrix{ \frac{n}{N-1} \cr r_i } u_i^{r_i} p^{-r_i}\,,\\
& = & \sum_{ \{ r_i \,: \sum_{i=1}^N r_i = n \} }
\Bigg\{  \prod_{i=1}^N
\pmatrix{ \frac{n}{N-1} \cr r_i } u_i^{r_i} \Bigg\}\,.
\end{eqnarray*}

\noindent Hence the result.

\endproof

\vskip 1cm

\noindent {\bf Example 2.2} Let $N=2\,.$ Then

\begin{eqnarray*}
Q^{(n)} & = & \sum_{r+s=n}\pmatrix{n \cr r}\pmatrix{n \cr s}u^r v^s \,,\\
& = & \sum_{r=0}^n \pmatrix{n \cr r}^2 u^r v^{n-r}\,.
\end{eqnarray*}

\noindent This was calculated, using a different approach in \cite{FS}.

\medskip

\endexample

\bigskip

These charges are conserved with respect to all the times, i.e.

\begin{equation}
\frac{\partial Q^{(m)}}{\partial t_n} =
\frac{\partial\Delta^{(m,n)}}{\partial x}
\label{eq:conlaw}
\end{equation}

\noindent for some function $\Delta^{(m,n)}\,.$ Before this function is
calculated it is first necessary to find the explicit form of the
evolution equations for the fields $u_i\,.$

\bigskip

\noindent {\bf Theorem 2.3}

\bigskip

The Lax equation (\ref{eq:lax}) implies the following evolution
equations for the fields

\[
u_{i,t_n} = A_i^{(n)} u_{i,x} + \sum_{j \neq i} u_i B_{ij}^{(n)} u_{j,x}
\]

\noindent where

\begin{equation}
A_i^{(n)}=\Big( \frac{n}{N-1}-1\Big)
\sum_{ \{ r_j\,: \sum_{j=1}^N r_j= n \} }
\Bigg[
\prod_{ \scriptstyle k=1 \atop \scriptstyle k\neq i}^N
\pmatrix{ \frac{n}{N-1} \cr r_k } u_k^{r_k}
\Bigg]
\pmatrix{ \frac{n}{N-1} -2 \cr r_i - 1} u_i^{r_i}
\label{eq:Ai}
\end{equation}
\noindent and
\begin{equation}
B_{ij}^{(n)} = \frac{n}{N-1} \sum_{ \{ r_j\,: \sum_{j=1}^N r_j= n-1 \} }
\Bigg[
\prod_{ \scriptstyle k=1 \atop \scriptstyle k\neq i,j}^N
\pmatrix{ \frac{n}{N-1} \cr r_k } u_k^{r_k}
\Bigg]
\pmatrix{ \frac{n}{N-1} -1 \cr r_i } u_i^{r_i}
\pmatrix{ \frac{n}{N-1} -1 \cr r_j } u_j^{r_j}\,.
\label{eq:Bij}
\end{equation}

\bigskip
\noindent {\bf Proof}

\bigskip

Since
\[
\L = \frac{1}{p} \prod_{i=1}^N (p+u_i)
\]
\noindent we have, on differentiating,

\begin{eqnarray*}
\frac{\partial\L}{\partial t_n} & = & \frac{1}{p} \sum_{j=0}^N
\Bigg[
\prod_{ \scriptstyle i=1 \atop \scriptstyle i \neq j}^N (p+u_i)
\Bigg]
u_{j,t_n}
\,, \\
& = & \L \sum_{i=1}^N \frac{u_{i,t_n}}{p+u_i} \,
\end{eqnarray*}
\noindent and similarly
\begin{eqnarray*}
\frac{\partial\L}{\partial x} & = &\L
\sum_{i=1}^N \frac{u_{i,x}}{p+u_i} \,,\\
\frac{\partial\L}{\partial p} & = &\L \Bigg\{-\frac{1}{p} +
\sum_{i=1}^N \frac{1}{p+u_i} \,\Bigg\}\,.
\end{eqnarray*}

\noindent The first of these derivatives implies that
\begin{eqnarray*}
u_{i,t_n} & = & \Res_{p=-u_i} \big[ \L^{-1} {\L_{t_n}}     \big] \,, \\
& = &  \Res_{p=-u_i} \big[ \L^{-1} \{ \M,\L \}  \big] \,,
\end{eqnarray*}

\noindent where, for convenience, we have defined $\M$ to be the
quantity $\M=(\L^{\frac{n}{N-1}})_{+}\,.$
The explicit form of $\M$ is easy
to calculate.

\begin{eqnarray*}
{\L}^{\frac{n}{N-1}} & = &
p^n \prod_{i=1}^N \sum_{r_i=0}^\infty \pmatrix{ \frac{n}{N-1}\cr r_i}
u_i^{r_i} p^{-r_i} \,, \\
& = & p^n \sum_{t=0}^\infty \sum_{ \{ r_i\,:\sum_{i=1}^N r_i=t\} }
\Bigg[
\prod_{i=1}^N \pmatrix{ \frac{n}{N-1}\cr r_i}
u_i^{r_i}
\Bigg]
\frac{1}{p^t}\,.
\end{eqnarray*}
\noindent So
\begin{eqnarray*}
{\M} & = & ({\L}^{\frac{n}{N-1}})_{+} \,, \\
& = & p^n \sum_{t=0}^n \sum_{ \{ r_i\,:\sum_{i=1}^N  r_i=t\} }
\Bigg[
\prod_{i=1}^N \pmatrix{ \frac{n}{N-1}\cr r_i}
u_i^{r_i}
\Bigg]
\frac{1}{p^t}\,.
\end{eqnarray*}

\bigskip

\noindent The derivatives of $\M$ are more complicated, but are easily
evaluated. Using these we obtain

\[
u_{i,t} = \Res_{p=-u_i} \sum_{t=0}^n p^{n-1}
\sum_{ \{ r_i\,:\sum_{i=1}^N r_i = t \} }
\Bigg[
\prod_{i=1}^N \pmatrix{ \frac{n}{N-1} \cr r_i} u_i^{r_i}
\Bigg]
\Bigg[
\sum_{i=1}^N (n-t) u_{i,x} \frac{1}{p+u_i} + \Big\{
	1-\sum_{k=1}^N \frac{1}{p+u_k} \Big\} \Bigg]
\]
\noindent and hence
\[
u_{i,t_n} = A_i^{(n)} u_{i,x} + \sum_{j\neq i} u_i B_{ij}^{(n)} u_{j,x}
\]

\noindent where

\begin{eqnarray*}
A_i^{(n)} & = & \sum_{t=0}^n (-1)^{n-t}
\sum_{ \{ r_j \,: \sum_{j=1}^N r_j = t \} }
\Bigg[
\prod_{\scriptstyle k=1 \atop \scriptstyle k\neq i}^N
\pmatrix{\frac{n}{N-1}\cr r_k} u_k^{r_k}
\Bigg]
\pmatrix{\frac{n}{N-1}\cr r_i} (n-t+r_i) u_i^{n+r_i-t} \,, \\
B_{ij}^{(n)} & = & \sum_{t=0}^n (-1)^{n-t}
\sum_{ \{ r_j \,: \sum_{j=1}^N r_j = t \} }
\Bigg[
\prod_{\scriptstyle k=1 \atop \scriptstyle k\neq i,j}^N
\pmatrix{\frac{n}{N-1}\cr r_k} u_k^{r_k}
\Bigg]
\pmatrix{\frac{n}{N-1}\cr r_i} 
\pmatrix{\frac{n}{N-1}\cr r_j} u_i^{n+r_i-t} u_j^{r_j-1}\,.
\end{eqnarray*}

\noindent To obtain the explicit forms (\ref{eq:Ai}) and (\ref{eq:Bij})
of these one has to perform a complicated resummation. In the
simplest $(N=2)$ case this amounts to using the identity
\[
\sum_{t=0}^n (-1)^{n-t} \sum_{r+s=t}
\pmatrix{n \cr r} \pmatrix{n \cr s} s u^{n+r-t} v^{s-1} =
n \sum_{r+s=n-1}
\pmatrix{n-1 \cr r} \pmatrix{n-1 \cr s} u^r v^s\,,
\]
\noindent the proof of which has  used various binomial identities.

\bigskip

\endproof

\bigskip

\noindent{{\bf Example 2.4} For general $N$ the first $(n=1)$
flow is

\[
u_{i,t}=-u_i u_{i,x} + {u_i \over N-1} \sum_{j=1}^N u_{j,x}
\]

\noindent and the $(N-1)^{\rm th}$ flow is:

\[
u_{i,t}=u_i \bigg(\prod_{j\neq i} u_i \bigg)_x
\]

\noindent For $N=2$ these are the same.

\bigskip

\noindent{{\bf Example 2.5} 
With $N=3$ and fields $u\,,v$ and $w$ the above formulae give:

\[
u_t = A u_x + u B v_x + u C w_x
\]

\noindent and cyclically, where 

\begin{eqnarray*}
A_n & = & (\frac{n}{2}-1)
\sum_{p+q+r=n}   \pmatrix{ \frac{n}{2}-1 \cr p-2 \cr }
                 \pmatrix{ \frac{n}{2}   \cr q   \cr }
                 \pmatrix{ \frac{n}{2}   \cr r   \cr } u^p v^q w^r \,, \\
& & \\
B_n & = & \frac{n}{2}
\sum_{p+q+r=n-1} \pmatrix{ \frac{n}{2}-1 \cr p   \cr }
                 \pmatrix{ \frac{n}{2}-1 \cr q   \cr }
                 \pmatrix{ \frac{n}{2}   \cr r   \cr } u^p v^q w^r \,, \\
& & \\
C_n & = & \frac{n}{2}
\sum_{p+q+r=n-1} \pmatrix{ \frac{n}{2}-1 \cr p   \cr }
                 \pmatrix{ \frac{n}{2}   \cr q   \cr }
                 \pmatrix{ \frac{n}{2}-1 \cr r   \cr } u^p v^q w^r \,.
\end{eqnarray*}

\bigskip

\noindent{\bf Example 2.6} A similar calculation may be performed for
the Benney hierarchy. In terms of the modified variables the first flow
is

\[
{\hat u}_{i,t} = {\hat u}_i \sum_{j \neq i} {\hat u}_{i,x}\,,
\]

\noindent this being the multicomponent
generalisation of (\ref{eq:3benney}).

\bigskip

\endexample

\bigskip

Having found the explicit form of the evolution equations one can now
derive the flux $\Delta^{(m,n)}$ corresponding to the conserved charge
(\ref{eq:charges},\ref{eq:charges2})\,.

\bigskip

\noindent {\bf Proposition 2.7}

\[
\Delta^{(m,n)} = \frac{N-1}{m+n}
\sum_{i=1}^N F_{u_i}^{(m)} \sum_{j \neq i} F_{u_j}^{(n)}
\]

\noindent where
\begin{eqnarray*}
F^{(m)} & = & \Delta^{(m,1)} \,, \\
& = & \sum_{ \{ r_k\,: \sum_{k=1}^N r_k= m+1 \} }
\Bigg[
\pmatrix{ \frac{m}{N-1} \cr r_k } u_k^{r_k}
\Bigg]\,.
\end{eqnarray*}

\bigskip

\noindent {\bf Proof}

\bigskip

It follows from the conservation law (\ref{eq:conlaw}) that

\[
\sum_{i=1}^N Q^{(n)}_{u_i} u_{i,t_m} =
\sum_{i=1}^N \Delta^{(n,m)}_{u_i} u_{i,x}
\]

\noindent and using the evolution equation this becomes, on
equating the various coefficients
of the derivatives,

\[
\sum_{i=1}^N u_i \Delta^{(m,n)}_{u_i} = \sum_{i=1}^N u_i Q^{(n)}_{u_i}
[A^{(m)}_i + \sum_{j \neq i} B^{(m)}_{ij} u_j]\,.
\]

\noindent It is clear, by power counting, that $\Delta^{(m,n)}$
must be a homogeneous
function of degree $(m+n)$ in the fields, so by Euler's theorem

\[
\Delta^{(m,n)} = \frac{1}{m+n}\sum_{i=1}^N u_i Q^{(n)}_{u_i}
[A^{(m)}_{i} + \sum_{j \neq i} B^{(m)}_{ij} u_j]\,.
\]

\noindent Since the $A_i$ and $B_{ij}$ are known it is now just a
matter of substituting
these into the above formula and simplifying the results. In fact
\begin{equation}
[A^{(m)}_{i}  +  \sum_{j \neq i} B^{(m)}_{ij} u_j] =
\sum_{ \{ r_k \,: \sum_{k=1}^N r_k = m\} }
m \Bigg[
\prod_{ \scriptstyle k=1 \atop \scriptstyle k \neq i}^N
\pmatrix{ \frac{m}{N-1} \cr r_k} u_k^{r_k}
\Bigg]
\pmatrix{ \frac{m}{N-1} - 1 \cr r_i} u_i^{r_i}
\label{eq:a1}
\end{equation}
\noindent the proof of which, once again, involves the use of
various binomial
identities. Let $F^{(m)} = \Delta^{(m,1)}\,,$ i.e. the flux
corresponding to
the first time flow. The above formulae simplify to give
\begin{equation}
F^{(m)} =  \sum_{ \{ r_k\,: \sum_{k=1}^N r_k= m+1 \} }
\Bigg[
\pmatrix{ \frac{m}{N-1} \cr r_k } u_k^{r_k}
\Bigg]\,,
\label{eq:flux}
\end{equation}
\noindent  so
\begin{eqnarray*}
F^{(m)}_{u_i} & = & \frac{1}{N-1} \Bigg[
\sum_{ \{ r_k \,: \sum_{k=1}^N r_k= m\} }
m \Bigg[
\prod_{ \scriptstyle k=1 \atop \scriptstyle k \neq i}^N
\pmatrix{ \frac{m}{N-1} \cr r_k} u_k^{r_k}
\Bigg]
\pmatrix{ \frac{m}{N-1} - 1 \cr r_i} u_i^{r_i}
\Bigg]\,,\\
& = & \frac{1}{N-1} 
[A^{(m)}_{i}  + \sum_{j \neq i} B^{(m)}_{ij} u_j]
\end{eqnarray*}
\noindent by using equations (\ref{eq:a1}). Thus
\[
\Delta^{(m,n)} = \frac{N-1}{m+n}
\sum_{i=1}^N  u_i Q^{(n)}_{u_i} F^{(m)}_{u_i}\,.
\]
\noindent The first order flow for general $N$ has
been calculated in an example
and it follows from this that
\[
F_{u_i}^{(n)} = \frac{2-N}{N-1} u_i Q_{u_i}^{(n)} +
\frac{1}{N-1} \sum_{j \neq i}
u_j Q_{u_j}^{(n)}\,.
\]
\noindent Inverting this gives
\[
u_i Q_{u_i}^{(n)} = \sum_{j \neq i} F_{u_j}^{(n)}
\]
\noindent and so
\begin{eqnarray*}
\Delta^{(m,n)} & = & \frac{N-1}{m+n}
\sum_{i=1}^N F_{u_i}^{(m)} \sum_{j \neq i} F_{u_j}^{(n)} \,, \\
& = & \frac{1}{m+n} \sum_{i\,,j} F^{(m)}_{u_i} G_{ij} F^{(n)}_{u_j}\,
\end{eqnarray*}
\noindent where $G_{ij}$ is the matrix with zero entries
along the diagonal and $N-1$
everywhere else.

\bigskip

\endproof

\bigskip

Note that $\Delta^{(m,n)}=\Delta^{(n,m)}\,,$ i.e.
\[
\frac{\partial Q^{(n)} }{\partial t_m} =
\frac{\partial Q^{(m)}}{\partial t_n}\,.
\]

\noindent It follows from this, together with the commutativity
of the flows,
that \cite{AoyamaKodama}

\[
\Delta^{(m,n)}=
\frac{\partial~}{\partial t_m}\frac{\partial~}{\partial t_n} F
\]

\noindent for some function $F\,.$ This is called the free
energy and is of great
importance in topological field theory \cite{Dubrovin,AoyamaKodama}.
There exist various algebraic relations between these
$\Delta^{(m,n)}\,,$ these
being derived from the differential Fay identities
\cite{TakasakiTakebe,CarrollKodama,Son}.
The matrix $G_{ij}$ is very closely connected to the
components of the metric ${\bf g}$ which defined the Hamiltonian
structure of
the hierarchy (see section 4). Where this is accidental or
indicative of some
deeper result is unknown. 

\bigskip

\section*{3. The algebraic structure of the charges}

\bigskip

Consider the conserved charges for the $N=2$ hierarchy
\[
Q^{(n)} = \sum_{r+s=n} \pmatrix{n \cr r} \pmatrix{n \cr s} u^r v^s\,.
\]
\noindent If one singles out one of the fields, say $u$,
this may be written as
\[
Q^{(n)} = u^n \sum_{r=0}^n \pmatrix{n \cr r}^2 \Big( \frac{v}{u} \Big)^r
\]
\noindent and so the properties of these charges can be
reduced to the study of
the polynomials
\[
f_n(x) = \sum_{r=0}^n \pmatrix{n \cr r}^2 x^r\,.
\]
\noindent These polynomials are examples of hypergeometric functions,
\[
f_n(x) = \, {}_2 F_{1} (-n,-n,1;x)
\]
\noindent (since $n$ is an integer this series will
automatically truncate leaving
a polynomial of degree $n$), and so the conserved charges may be
written as
\[
Q^{(n)} = u^n {}_2 F_{1}\Big(-n,-n,1;\frac{v}{u} \Big)\,.
\]
\noindent Using the identities
\begin{eqnarray*}
{}_2 F_{1}(a,b,c;z) & = & (1-z)^{-b} {}_2 F_{1}
\Big(c-a,b,c;\frac{z}{z-1}\Big)\,, \\
P_n(z) & = & {}_2 F_{1} \Big(n+1,-n,1;\frac{1-z}{2} \Big)
\end{eqnarray*}
\noindent (where $P_n(z)$ is the Legendre polynomial of degree $n$)
one obtains
\[
Q^{(n)} = (u-v)^n P_n \Big( \frac{u+v}{u-v} \Big)\,,
\]
\noindent and so a generating function for these charges may be
constructed using
the well known generating function for Legendre polynomials.
Such a generating
function was obtained in \cite{FS} using a different
approach based on the construction of
recursion relations.

\bigskip

In the multicomponent case one may perform a similar calculation and
express the
charges in terms of generalized hypergeometric functions \cite{Gelfand}.
Defining new
variables $x_i$ by
\[
x_i = \cases{ u_N \,, & $\quad\quad\quad i=N\,,$ \cr \frac{u_i}{u_N} \,, &
$\quad\quad\quad i\neq N$}
\]
\noindent equation (\ref{eq:charges2}) may be written as
\begin{eqnarray*}
Q^{(n)} & = & \Bigg[\Big( \frac{n}{N-1} \Big)! \Bigg]^N
 x_N^n \sum_{ {\bf r} \in {\bf Z}^{N-1}} {x^r \over
\prod_{k=1}^{2N} \Gamma(1+\mu_k({\bf r}) + \gamma_k ) } \,, \\
& = & \Big( \frac{n}{N-1} \Big)^N
 x_N^n \,\, {\bf F}(\mu,\gamma;{\bf x})
\end{eqnarray*} 

\noindent where, for convenience, $x^r$ is defined to be
\[
x^r=x_1^{r_1} \ldots x_{N-1}^{r_{N-1}}\,,
\]
\noindent and $r_i$ are the components of the vector ${\bf r}\,.$ The
linear map
$\mu\,: {\bf Z}^{N-1} \rightarrow {\bf Z}^{2N}$ is defined by

\[
\mu_k({\bf r}) = \sum_{j=1}^{N-1} a_{kj} r_j
\]

\noindent where

\[
a_{kj}=\pmatrix{
+1 & 0 & \ldots & 0 \cr
0 & +1 & \ldots & 0 \cr
\vdots& \vdots &  & \vdots \cr
0 & 0 & \ldots & +1 \cr
+1 & +1 & \ldots & +1 \cr
-1 & 0 & \ldots & 0 \cr
0 & -1 & \ldots & 0 \cr
\vdots& \vdots &  & \vdots \cr
0 & 0 & \ldots & -1 \cr
-1 & -1 & \ldots & -1 \cr}\,
\]

\noindent and the constant vector $\gamma$ is
\[
\gamma=\pmatrix{
0 \cr
0 \cr
\vdots \cr
n ( {2-N\over N-1} ) \cr
{n \over N-1} \cr
{n \over N-1} \cr
\vdots \cr
n \cr}
\]

\noindent These manipulations are a manifestation of the Ore-Sato
Theorem for 
generalized hypergeometric functions. Thus the charges are defined
in terms of a
linear map $\mu$ and a constant vector $\gamma\,,$ or in terms of a
lattice ${\bf B}$ inside
${\bf Z}^{2N}$ as the image of ${\bf Z}^{N-1}$ under the linear map
$\mu\,.$
These generalized hypergeometric functions a related (but not identical)
to those
studied by Gelfand et al.\cite{Gelfand}.

\bigskip

The fluxes (\ref{eq:flux}) also have a similar description,
\[
F^{(n)} = \Bigg[\Big( \frac{n}{N-1} \Big)!\Bigg]^N x^n_N \,\,
{\bf F}(\mu,\gamma^\prime;{\bf x})
\]
\noindent where the linear map $\mu$ is the same but the constant vector
has
changed to
\[
\gamma^\prime = \gamma +\pmatrix{
0 \cr
0 \cr
\vdots \cr
1 \cr
0 \cr
0 \cr
\vdots \cr
1 \cr}\,.
\]
\noindent So both the charges and the flux are defined in terms of
the same lattice
${\bf B} \in {\bf Z}^{2N}\,,$ the difference being in the constant
vector. One
remaining problem is to understand the structure of the more general
fluxes
$\Delta^{(m,n)}$ in terms of these structures.

\bigskip

\section*{4. The Hamiltonian structure of the hierarchy}

The evolution equations studied in this paper are specific examples of
hydrodynamic
equations, i.e. they are of the form

\begin{equation}
u^i_t = \sum_j V^i_j({\bf u}) u^j_x\,,
\label{eq:hydro}
\end{equation}

\noindent the simplest non-trivial example being the Monge equation
$u_t=u u_x\,.$
One must, in general, take care with the indices on the fields as they
transform
as tensors under an arbitrary transformation
$u_i\rightarrow {\tilde u}_i({\bf u})\,.$
However, to avoid expression such as $(u^i)^{r_i}$ all indices have been
lowered for notational convenience. There is an extensive literature
on such
hydrodynamic equations (recently they have attracted much interest from
their role
in topological field theories). One important result concerns the
Hamiltonian
structure of these systems.

\bigskip

A system (\ref{eq:hydro}) of hydrodynamic type is said to be Hamiltonian
if
there exists a Hamiltonian $H=\int dx \, h({\bf u})$ and a
Hamiltonian operator

\begin{equation}
{\hat A}^{ij} = g^{ij} ({\bf u}) {d\phantom{x}\over dx} +
b^{ij}_{\phantom{ij}k} ({\bf u}) u^k_x
\label{eq:hamstructure}
\end{equation}

\noindent which defines a skew-symmetric Poisson bracket on functionals

\[
\{I,J\} = \int dx \, {\delta I\over \delta u^i(x) } {\hat A}^{ij}
                     {\delta J\over \delta u^j(x) }
\]

\noindent which satisfies the Jacobi identity and which generates the
system

\[
u^i_t = \{ u^i(x), H\}\,.
\]

\noindent Dubrovin and Novikov \cite{DubrovinNovikov} proved necessary and
sufficient
conditions for
${\hat A}^{ij}$ to be a Hamiltonian operator in the case when $g^{ij}$
is not
degenerate. These are:

\bigskip

a) ${\bf g}=(g^{ij})^{-1}$ defines a Riemannian metric,

\medskip

b) $b^{ij}_{\phantom{ij}k} = - g^{is} \Gamma^{j}_{\phantom{j}sk}\,,$ where
$\Gamma^{j}_{\phantom{j}sk}$ is the Christoffel symbol generated by
${\bf g}\,,$

\medskip

c) the Riemann curvature tensor of $\bf g$ vanishes.

\bigskip

\noindent The system (\ref{eq:hydro}) may then be written as

\begin{equation}
u^i_t = ( g^{is} \nabla_s \nabla_j h ) u^j_x  \label{eq:hamhydro}
\end{equation}

\noindent where $\nabla$ is the covariant derivative generated by
${\bf g}\,.$
Expanding these in terms of the Christoffel symbols gives

\begin{eqnarray*}
u^i_t & = & [g^{ij} \partial_k \partial_j h - g^{is}
\Gamma^j_{sk}\partial_j h]
u^k_x \,, \\
& = & {\widetilde A}_i u^i_x + \sum_{k \neq i} u^i
{\widetilde B}_{ik} u^k_x \quad\quad\quad\quad (\rm no~sum)
\end{eqnarray*}

\noindent where

\begin{eqnarray}
{\widetilde A}^i & = & \sum_j g^{ij} \partial_i \partial_j h -
\sum_{s\,,j} g^{is}
\Gamma^j_{si} \partial_j h\,,
\label{eq:tildeA} \\
{\widetilde B}_{ik} & = & \sum_j g^{ij} \partial_k \partial_j h -
\sum_{s\,,j} g^{is}
\Gamma^j_{sk} \partial_j h \quad (k\neq i) \,,
\label{eq:tildeB}
\end{eqnarray}

\noindent Thus to find the Hamiltonian structure for the hierarchy
(\ref{eq:lax})
one needs to find the metric and the Hamiltonians.

\bigskip

One approach is to first diagonalise the system, there being the
following simple formulae
for the metric coefficients for diagonal systems \cite{Tsarev}:

\[
{\partial_i V_j \over (V_i - V_j)} = \frac{1}{2} \partial_i \log g_{jj}
\]

\noindent where $V^i_j ({\bf u}) = V_j({\bf u}) \delta^i_j$ (no sum). For
$N=2$ such an approach was adopted in \cite{FS}. However, to find the
corresponding
transformation for arbitrary $N$ is difficult and another approach is
needed,
the $N=2$ result suggesting the form of the general result.

\bigskip

\noindent {\bf Theorem 4.1}

\bigskip

The Hamiltonian structure for the hierarchy (\ref{eq:lax}) is given by
(\ref{eq:hamstructure}), where the zero-curvature
metric is given by

\begin{equation}
{\bf g} = \frac{1}{N-1} \sum_{i \neq j} \frac{du_i}{u_i} \,
\frac{du_j}{u_j}
\label{eq:hammetric}
\end{equation}

\noindent and the Hamiltonians generating these flows are given by

\[
h^{(n)} = \frac{1}{n} Q^{(n)} \,,
\]
\noindent where the $Q^{(n)}$ are conserved charges given by
({\ref{eq:charges},\ref{eq:charges2})\,.

\bigskip

\noindent {\bf Proof}

\bigskip

This ${\bf g}$ defines a non-degenerate Riemannian metric, and by
changing coordinates to
${\tilde u}_i=\log(u_i)$ it is clear that the metric has vanishing
Riemannian
curvature tensor. The components of the inverse metric are

\begin{eqnarray*}
g^{ii} = &  -(N-2) & u_i u_i \quad\quad (\rm no~sum) \,, \\
g^{ij} = &  & u_i u_j  \quad\quad (i \neq j)
\end{eqnarray*}

The only non-zero Christoffel symbols are $\Gamma^i_{ii}$ (no sum).
This is easily
proved by noting that in the ${\tilde u}_i$ coordinates the
Christoffel symbols are
all zero. The Christoffel symbols can then be calculated using
transformation law
for these symbols.
This yields

\[
\Gamma^i_{ii} = - \frac{1}{u_i} \quad (\rm no~sum)
\]

\noindent with all other components being zero.
Thus, by the theorem of Dubrovin and Novikov, this metric defines a
Hamiltonian
structure. To show that this is the dispersionless Toda hierarchy one
has to verify
that the $\widetilde A_i$ given by (\ref{eq:tildeA}) and the
$\widetilde B_{ij}$ given by
(\ref{eq:tildeB}) and $h^{(n)}=\frac{1}{n} Q^{(n)}$ are the same as
given by
equations ({\ref{eq:Ai}) and ({\ref{eq:Bij})
in theorem 2.3.

\bigskip

\begin{eqnarray*}
{\widetilde A}^i & = & \sum_j g^{ij} \partial_i \partial_j h -
\sum_{s\,,j} g^{is}
\Gamma^j_{si} \partial_j h\,,\\
& = & \Big[ g^{ii} \partial_i \partial_i h + g^{ii}
\frac{1}{u_i}\partial_i h \Big] +
\sum_{j \neq i} g^{ij} \partial_i \partial_j h\,.
\end{eqnarray*}

\noindent Now
\[
h^{(n)} = \frac{1}{n} \sum_{ \{ r_i \,: \sum_{i=1}^N r_i = n \} }
\Bigg\{  \prod_{i=1}^N
\pmatrix{ \frac{n}{N-1} \cr r_i } u_i^{r_i} \Bigg\}\,,
\]
\noindent and substitution of this into the above gives

\begin{eqnarray*}
{\widetilde A}_i^{(n)} & = & \Big( \frac{n}{N-1}-1\Big)
\sum_{ \{ r_j\,: \sum_{j=1}^N r_j= n \} }
\Bigg[
\prod_{ \scriptstyle k=1 \atop \scriptstyle k\neq i}^N
\pmatrix{ \frac{n}{N-1} \cr r_k } u_k^{r_k}
\Bigg]
\pmatrix{ \frac{n}{N-1} -2 \cr r_i - 1} u_i^{r_i} \,, \\
& = & A_i^{(n)}\,.
\end{eqnarray*}

\noindent The proof that ${\widetilde B}_{ij}^{(n)} = B_{ij}^{(n)}$ is
similar,
both using various
binomial identities. Hence the result.

\bigskip

\endproof

\bigskip

This theorem thus determines the Hamiltonian structure of the
dispersionless Toda hierarchy.

\bigskip

\section*{5. Miscellany}

\bigskip

This section contains some miscellaneous results on the multicomponent
Toda
hierarchy.

\bigskip

\subsection*{5.1. Reductions}

\bigskip

Besides the discrete symmetry $u_i \mapsto u_{\sigma(i)}\,,$ where
$\sigma \in S_N$ is an element of the permutation group,
the multicomponent
Toda hierarchy possesses many other symmetries, as befits an integrable
system. For example, one has the continuous symmetries corresponding to
time and space translations, and various scaling symmetries. These may
be used to derive examples of integrable dynamical systems.

\bigskip

With the ansatz

\[
u_i(x,t) = (n x)^\frac{1}{n} z_i(t)
\]

\noindent for the $n^{\rm th}$ flow (this being a particular example
of a scaling symmetry) the $x$-dependence cancels and one is left with
a dynamical system for the $z_i$ variables.

\bigskip

\noindent {\bf Example 5.1} With $n=1$ the above ansatz reduces the system
(\ref{eq:3benney}) to the dynamical system

\begin{eqnarray*}
{\dot z}_1 & = &  z_1 (z_2 + z_3) \,, \\
{\dot z}_2 & = &  z_2 (z_3 + z_1) \,, \\
{\dot z}_3 & = &  z_3 (z_1 + z_2) \,.
\end{eqnarray*}

\noindent This system has been studied and integrated by Bureau
\cite{Bureau}.

\bigskip

\endexample

\bigskip

In general, such dynamical system are of the form

\[
{\dot z}_i = z_i \, ({\rm polynomial~in~the~}z_j{\rm ~variables})
\]

\noindent (since $A_i^{(n)}$ given by equation (\ref{eq:Ai}) always
has $u_i$ as a factor), which suggest on should introduce variables
$\phi_i = \log z_i\,.$ The resultant dynamical systems are then in the
so-called Lotka-Volterra form, the nonlinear terms being sums of
products of exponentials.

\bigskip

As mentioned in section 2, one may obtain other systems by a
conflation of the fields (or equivalently, by having multiple
roots in the equation ${\cal L}(p)=0$). As an example, the
system (\ref{eq:3example})

\begin{eqnarray*}
u_t & = & \frac{1}{2} u ( -u_x+v_x+w_x ) \,,\\
v_t & = & \frac{1}{2} v ( +u_x-v_x+w_x ) \,,\\
w_t & = & \frac{1}{2} w ( +u_x+v_x-w_x ) \,.
\end{eqnarray*}

\noindent becomes, one setting $v=w\,,$

\begin{eqnarray*}
u_t & = & \frac{1}{2} u (2 v_x - u_x) \,, \\
v_t & = & \frac{1}{2} v u_x
\end{eqnarray*}

\noindent and finally, on setting $u=v\,,$

\[
u_t = \frac{1}{2} u u_x\,.
\]

\noindent The Hamiltonian structure of these later systems
is obtained by restricting the metric (which defined the Hamiltonian
structure of the original system) in the first case to the surface
$v=w$ and in the second case to the line $u=v=w\,.$ Schematically
these systems may be written as $\{1,1,1\} \,, \{2,1\}$ and $\{3\}$
respectively to show the confluence of the fields (the discrete
permutation symmetry ensuring that $\{2,1\}$ is the same as $\{1,2\}$),
and hence the possible reductions may be summarised in the diagram

\[
\{1,1,1\} \longrightarrow \{2,1\} \longrightarrow \{3\}\,.
\]

\noindent With more fields the are many more possible reductions
and a simple combinatorial argument gives the number of such
subsystems (including the original system) to be
$\displaystyle{\frac{2N^2+7 +(-1)^N}{8}}$ for the $N$-component Toda
hierarchy.
Figure 1 shows the possible reductions for $N=4$ and $N=5\,.$

\bigskip

\begin{figure*}
\begin{picture}(250,250)(-50,0)
\put(50,200){\{1,1,1,1\}}
\put(72,192){\vector(0,-4){20}}
\put(55,160){\{2,1,1\}}
\put(72,152){\vector(1,-1){20}}
\put(72,152){\vector(-1,-1){20}}
\put(30,120){\{3,1\}}
\put(89,120){\{2,2\}}
\put(52,112){\vector(1,-1){20}}
\put(92,112){\vector(-1,-1){20}}
\put(64,79){\{4\}}
\put(175,200){\{1,1,1,1,1\}}
\put(202,192){\vector(0,-4){20}}
\put(180,160){\{2,1,1,1\}}
\put(202,152){\vector(1,-1){20}}
\put(202,152){\vector(-1,-1){20}}
\put(152,120){\{2,2,1\}}
\put(219,120){\{3,1,1\}}
\put(182,112){\vector(0,-4){40}}
\put(222,112){\vector(0,-4){40}}
\put(182,112){\vector(1,-1){40}}
\put(222,112){\vector(-1,-1){40}}
\put(160,60){\{4,1\}}
\put(219,60){\{3,2\}}
\put(182,52){\vector(1,-1){20}}
\put(222,52){\vector(-1,-1){20}}
\put(194,19){\{5\}}
\end{picture}
\caption{Possible reductions of the $N=4$ and $N=5$ Toda hierarchies}
\end{figure*}
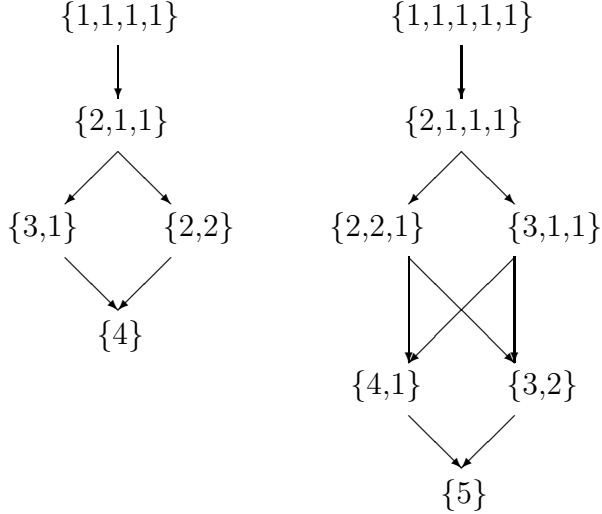

\bigskip

\subsection*{5.2. Special flows}

\bigskip

It is clear from the Lax equation (\ref{eq:lax}) and the integral for the
conserved charges (\ref{eq:charges}) that if $n=(N-1)m$ for some integer
$m$ then some simplification might
occur, since for such a value of $n$ the Lax equation

\[
\frac{\partial{\cal L}}{\partial {\hat t}_m} = \{ ( {\cal L}^m)_{+}, \L \}
\]

\noindent (where ${\hat t}_m = t_{(N-1)m}\,$) and the integral
form for the conserved
charges

\[
{\hat Q}^{(m)} = \frac{1}{2\pi i} \oint \L^m \frac{dp}{p}
\]

\noindent do not involve fractional powers of the Lax function.
These flows are
best studied by a change of variables

\[
{\hat u}_i = \prod_{k \neq i} u_i\,,
\]

\noindent this being the generalisation of the one used in example 1.3.
In these variables the evolution equations for these flows
take the form

\begin{equation}
{\hat u}_{i,{\hat t}_m} = {\hat A}_i^{(m)} {\hat u}_{i,x} +
\sum_{j \neq i} {\hat u}_i {\hat B}_{ij}^{(m)} {\hat u}_{j,x}
\label{eq:hatevol}
\end{equation}

\noindent where

\begin{eqnarray*}
{\hat A}_i^{(m)} & = & (m-1) \sum_{ \{ r_j\,: \sum_{j=1}^N r_j= m \} }
\Bigg[
\prod_{ \scriptstyle k=1 \atop \scriptstyle k\neq i}^N
\pmatrix{ m \cr r_k } {\hat u}_k^{r_k}
\Bigg]
\pmatrix{ m -2 \cr r_i - 1} {\hat u}_i^{r_i} \,, \\
{\hat B}_{ij}^{(m)} & = & m \sum_{ \{ r_j\,: \sum_{j=1}^N r_j= m-1 \} }
\Bigg[
\prod_{ \scriptstyle k=1 \atop \scriptstyle k\neq i,j}^N
\pmatrix{ m \cr r_k } {\hat u}_k^{r_k}
\Bigg]
\pmatrix{ m -1 \cr r_i } {\hat u}_i^{r_i}
\pmatrix{ m -1 \cr r_j } {\hat u}_j^{r_j}\,.
\end{eqnarray*}

\noindent and the corresponding conserved charges are

\[
{\hat Q}^{(m)} = \sum_{ \{ r_i\,: \sum_{j=1}^N r_j = m\} }
\Bigg[
\prod_{i=1}^N \pmatrix{m \cr r_i } {\hat u}_i^{r_i}
\Bigg]\,.
\]

\noindent Superficially these equations look very similar to their
unhatted counterparts. However, it is important to note that
$N$ only appears in a subordinate role, i.e. instead of the
summation being over the set $\{ r_i\,: \sum r_i = (N-1)m\}$
the change of variable enables this to be
written as a summation over the
set $\{ r_i\,: \sum r_i = m\}$ irrespective of the value of $N\,.$
Similarly $N$ does not enter the binomial coefficients
that appear in the above
formulae. One consequence of this is, if one denotes the evolution
equation (\ref{eq:hatevol}) by $\Gamma(N,m)\,,$ then a possible
reduction is to set ${\hat u}_N=0$ and the resultant system is just
the $(N-1)$-component system, i.e.

\[
\Gamma(N,m) \bigg\vert_{{\hat u}_N=0} = \Gamma(N-1,m)
\]

\noindent for all values of $m\,.$

\bigskip

The Hamiltonian structure of this system is easily derived by
performing the change of variable on the metric (\ref{eq:hammetric}),
the result being

\[
{\bf {\hat g}} = \frac{1}{(N-1)^2} \Bigg[ (2-N) \sum_{i=1}^N
\Big( \frac{d{\hat u}_i}{{\hat u}_i} \Big)^2 + \sum_{r \neq s}
\frac{d{\hat u}_r}{{\hat u}_r}\,
\frac{d{\hat u}_s}{{\hat u}_s} \Bigg]
\]

\noindent (actually, the inverse metric

\[
{\hat {\bf g}}^{ij} =
\cases{ (N-1) {\hat u}_i {\hat u}_j & $i \neq j \,,$ \cr
0 & otherwise,}
\]

\noindent is simplier).

\bigskip

These calculations are not, perhaps, very illuminating, and the motivation
for this subsection is best seen by means of the following examples.

\bigskip

\noindent{\bf Example 5.2} For general $N$ the $(N-1)^{\rm th}$ flow
(i.e. $m=1$)
has been
calculated in example 2.4, namely,

\[
u_{i,t} = u_i \bigg( \prod_{j \neq i} u_i \bigg)_x\,.
\]

\noindent In the hatted variables these simplify to

\[
{\hat u}_i = {\hat u}_i \sum_{j \neq i} {\hat u}_{j,x}\,.
\]

\noindent This is the first flow of the Benney hierarchy (see example 2.6)
written in terms of the modified variables.
Note that this contains only quadratic nonlinearities irrespective
of the value of $N\,,$ unlike its unhatted counterpart whose nonlinearities
depend on the value of $N\,.$

\bigskip

\noindent{\bf Example 5.3} For $N=3$ the Lax equation (\ref{eq:lax})
gives the
evolution equations for $n=2$

\[
u_{t_2} = u w v_x + u v w_x  \quad\quad(\rm and~cyclically)
\]

\noindent and for $n=4$

\[
u_{t_4} = 2 uvw(v+w)u_x+2uw(2uv+uw+vw)v_x+2uv(uv+2uw+vw)w_x
\quad(\rm and~cyclically)
\]

\noindent together with the conserved charges

\begin{eqnarray*}
Q^{(2)} & = & uv+vw+wu \,, \\
Q^{(4)} & = & u^2 v^2 + v^2 w^2 + w^2 u^2 + 4 uvw(u+v+w)\,.
\end{eqnarray*}

\noindent In terms of the new hatted variables ${\hat u} = v w $ etc.
these flows
become

\begin{eqnarray*}
{\hat u}_{{\hat t}_1} & = & {\hat u} ({\hat v}_x + {\hat w}_x) \,,\\
{\hat u}_{{\hat t}_2} & = & (2{\hat u}{\hat v} + 2
{\hat u}{\hat w}){\hat u}_x +
(2{\hat u}^2 + 2 {\hat u}{\hat v} + 4 {\hat u}{\hat w}) {\hat v}_x +
(2{\hat u}^2 + 4 {\hat u}{\hat v} + 2 {\hat u}{\hat w}) {\hat w}_x
\end{eqnarray*}

\noindent (together with their cyclic permutations) and the
conserved charges
become

\begin{eqnarray*}
{\hat Q}^{(1)} & = & {\hat u}+{\hat v}+{\hat w} \,, \\
{\hat Q}^{(2)} & = & {\hat u}^2 + {\hat v}^2 + {\hat w}^2 + 4
({\hat u}{\hat v}+
{\hat v}{\hat w} + {\hat w}{\hat u})\,.
\end{eqnarray*}

\noindent These flows are thus simpler in the hatted variables.
Note that it is
now possible to set, for example, ${\hat w}=0$ and these formulae
reduce to
those for $N=2$ (for which there is no distinction between hatted and
unhatted
variables).

\bigskip

\endexample

\bigskip

\section*{6. Rational Lax functions}

\bigskip

The Lax function so far studied in this paper has been,
apart from an overall
$p^{-1}\,,$ a polynomial function. However, the Lax formalism is
valid for much
more general functions such as rational functions:

\begin{equation}
\L = { \prod_{i=1}^N (p+u_i) \over \prod_{i=N+1}^{N+M} (p+u_i)} \,,
\label{eq:rationalLax}
\end{equation}

\noindent or

\[
\L = \prod_{i=1}^{M+N} (p+u_i)^{\epsilon_i}
\]

\noindent provided $N>M\,.$ Here $\epsilon_i=\pm 1$ but this can
be extended to cope
with repeated roots and zeros. The $N>M$ condition enables this
rational Lax function
to be written as
\[
\L ={\rm polynomial~of~degree~}(N-M) ~ +
\sum_{i=N+1}^{N+M} {\rm simple~poles}\,.
\]
\noindent It should be clear from the earlier proofs that they
may be extended
to such rational potentials with little change. The results in
this section will
therefore be given without proof.

\bigskip

The conserved charges, now defined by the integral formula

\[
Q^{(n)} = \frac{1}{2\pi i} \oint \L^{ \frac{n}{N-M} } \frac{dp}{p}\,,
\]

\noindent are

\[
Q^{(n)} = \sum_{ \{ r_i \,: \sum_{i=1}^{N+M} r_i= n \} }
\Bigg\{  \prod_{i=1}^{N+M}
\pmatrix{ \frac{\epsilon_i n}{N-M} \cr r_i } u_i^{r_i} \Bigg\}\,,
\]

\noindent and the evolution of the fields, given by the Lax equation

\[
\frac{\partial\L}{\partial t_n} = \{ (\L^{\frac{n}{N-M}})_{+} , \L \}\,,
\]

\noindent are

\[
u_{i,t_n} = A_i^{(n)} u_{i,x} + \sum_{j \neq i} u_i B_{ij}^{(n)} u_{j,x}
\]

\noindent where

\[
A_i^{(n)}=\Big( \frac{\epsilon_i n}{N-M}-1\Big)
\sum_{ \{ r_j\,: \sum_{j=1}^{N+M} r_j= n \} }
\Bigg[
\prod_{ \scriptstyle k=1 \atop \scriptstyle k\neq i}^{N+M}
\pmatrix{ \frac{\epsilon_k n}{N-M} \cr r_k } u_k^{r_k}
\Bigg]
\pmatrix{ \frac{\epsilon_i n}{N-M} -2 \cr r_i - 1} u_i^{r_i}
\]

\noindent and

\[
B_{ij}^{(n)} = \frac{\epsilon_j n}{N-M}
\sum_{ \{ r_j\,: \sum_{j=1}^{N+M} r_j= n-1 \} }
\Bigg[
\prod_{ \scriptstyle k=1 \atop \scriptstyle k\neq i,j}^{N+M}
\pmatrix{ \frac{\epsilon_k n}{N-M} \cr r_k } u_k^{r_k}
\Bigg]
\pmatrix{ \frac{\epsilon_i n}{N-M} -1 \cr r_i } u_i^{r_i}
\pmatrix{ \frac{\epsilon_j n}{N-M} -1 \cr r_j } u_j^{r_j}\,.
\]

\bigskip

\noindent{\bf Example 6.1} The simplest case is $N=2\,,M=1\,,$ and the
Lax function will be written as

\[
\L=\frac{ (p+u)(p+v) }{p+w}\,.
\]

\noindent Hence a consistent reduction is to set $w=0\,,$
with which the results
simplify further. The conserved charges are 

\[
Q^{(n)} = \sum_{r+s+t=n} \pmatrix{n \cr r} \pmatrix{n \cr s}
\pmatrix{-n \cr ~t}
u^r v^s w^t\,.
\]

\noindent The binomial coefficients have to be interpreted formally,
as indicated
in section 1. For example, the first two charges are

\begin{eqnarray*}
Q^{(1)} & = & u+v - w \,, \\
Q^{(2)} & = & u^2 + v^2 + 3 w^2 + 4 (uv - vw - uw)\,,
\end{eqnarray*}

\noindent and the first two flows are given by

\[
\pmatrix{u \cr v\cr w\cr}_{t_1} =
\pmatrix{ 0 & u & -u \cr v & 0 & -v \cr w & w & -2w}
\pmatrix{u \cr v\cr w\cr}_x
\]

\noindent and

\[
\pmatrix{u \cr v\cr w\cr}_{t_2} =
\pmatrix{ 2u(v-w) & 2u(u+v-2w) & -2u(u+2v-3w)\cr
          2v(u+v-2w) & 2v(u-w) & -2v(2u+v-3w)\cr
          2w(u+2v-3v) & 2w(2u+2v-3w) & -6w(u+v-2w) }
\pmatrix{u \cr v\cr w\cr}_x\,.
\]

\noindent Once again, care has to be taken in evaluating the
binomial coefficients.

\bigskip

\noindent{\bf Example 6.2} Let $M=N-1$ and $u_i=0$ for
$i=N+1\,,\ldots \,,N+M\,.$
With these the Lax function takes the form

\begin{eqnarray*}
\L & = & \frac{ \prod_{i=1}^N (p+u_i) }{p^{N-1}} \,, \\
& = & p \prod_{i=1}^N \bigg( 1 + \frac{u_i}{p} \bigg) \,, \\
& = & p + \sum_{r=0}^N \frac{S_r}{p^r} \,, \\
& = & \L_B \,.
\end{eqnarray*}

\noindent Thus the Lax function $\L_B$ for the Benney hierarchy is a
special
case of the rational Lax function. Thus one obtains an explicit
family of conservation
laws and evolution equations for the Benney hierarchy.

\bigskip

\endexample

\bigskip

In fact, the rational Lax function (\ref{eq:rationalLax}) itself
may be considered
as a reduction of the infinite component Lax function

\[
\L_\infty = p^{N^\prime} + \sum_{n=-\infty}^{N^\prime -1} S_i p^n\,.
\]

\noindent By formally expanding (\ref{eq:rationalLax}) one
obtains the above
Lax function, with various conditions of the fields. This and
some other reductions
of this Lax function, and the resulting systems, have been studied by
\cite{GibbonsTsarev}.

\section*{7. Logarithmic charges}

\bigskip

In section $4$ it was shown that the conserved charges (for $N=2$)
may be written in terms of
Legendre polynomials
\begin{equation}
Q^{(n)} = (u-v)^n \, P_n \Big( \frac{u+v}{u-v} \Big)\,.
\label{eq:uvcharges}
\end{equation}
\noindent This suggests that one should introduce the new variables
\begin{eqnarray*}
r & = & u-v \,, \\
\cos\theta & = & \frac{u+v}{u-v}\,.
\end{eqnarray*}
\noindent In these new variables the differential equation for the
conserved
charges
\[
(u Q_u)_u = (v Q_v )_v
\]
\noindent becomes the axially symmetric Laplace equation,
and so one obtains the
conserved charges (\ref{eq:uvcharges}). However there is another
family of charges
given by Legendre functions of the second kind, the first few conservation
laws begin:
\begin{eqnarray*}
\Big[ \frac{1}{2} \log(uv) \Big]_t & = &
\Big[ \frac{1}{2} (u+v) \Big]_x \,,\\
\Big[ \frac{1}{2} (u+v)\log(uv) \Big]_t & = &
\Big[ \frac{1}{2} uv \log(uv) +\frac{1}{4} (u^2+v^2) \Big]_x\,.
\end{eqnarray*}

\bigskip

This raises the question of how these charges are related to the
Lax formalism.
The answer is that they are given by the formula
\begin{equation}
{\widetilde Q}^{(n)} = \frac{1}{2\pi i} \oint \L^n (\log\L - c_n)
\frac{dp}{p}
\label{eq:logcharges}
\end{equation}
\noindent where
\[
c_n = \sum_{j=1}^n \frac{1}{j}\,, \quad\quad c_0 = 0 \,.
\]

\noindent This results in the scaling symmetry
\cite{EguchiYang,EguchiYangHori}

\[
\frac{\partial~}{\partial\L} \big[ \L^n(\log\L -c_n)\big] =
n \big[ \L^{n-1} (\log\L - c_{n-1})\big]\,,
\]

\noindent since $c_n = c_{n-1} + 1/n\,.$
Once again using the modified variables this integral may be performed
for all values of $n$ and $N\,.$ For simplicity we restrict
attention to the $N=2$
case. Conceptually the general case is the same, just notationally
more complex.

\bigskip

\noindent {\bf Theorem 7.1}

\bigskip

The logarithmic charges defined by (\ref{eq:logcharges}) are

\begin{equation}
{\widetilde Q}^{(n)} = \frac{1}{2} \log(uv) \sum_{r+s=n}
\pmatrix{n\cr r} \pmatrix{n\cr s} u^r v^s-
\sum_{r+2=n} \pmatrix{n \cr r} \pmatrix{n \cr s} c(r,s) u^r v^s
\label{eq:logcharges2}
\end{equation}

\noindent where
\[
c(r,s) = \sum_{j=1}^r \frac{1}{j} \,\, - \sum_{j=s+1}^{r+s} \frac{1}{j}
\]

\noindent and $c(0,r)=0$ (so $c(r,s)=c(s,r)$ for all $r$ and $s$).

\bigskip

\noindent{\bf Proof}

\bigskip 

Care has to be taken with the interpretation of $\log\L$ to avoid
$\log p$-terms.
Since
\[
\log\L = +\log p + \log\Big( 1+ \frac{u}{p} \Big) +
\log\Big( 1+ \frac{v}{p} \Big)
\]
\noindent and
\[
\log\L = -\log p + \log\Big( 1 + \frac{p}{u}\Big) +
\log\Big( 1+ \frac{p}{v} \Big) +\log(uv)
\]
\noindent then
\[
\log\L=\frac{1}{2} \Bigg[
\log\Big( 1+ \frac{u}{p} \Big) + \log\Big( 1+ \frac{v}{p} \Big)+\log(uv)+
\log\Big( 1 + \frac{p}{u}\Big) + \log\Big( 1+ \frac{p}{v} \Big) \Bigg]\,.
\]

\bigskip

To evaluate the integral one Taylor expands the function as power
series in $p$ and
uses the residue theorem. Since
\[
\L^n \log\Big( 1+ \frac{u}{p} \Big) = p^n
\sum_{r=0}^n \sum_{s=0}^n \sum_{t=0}^\infty \pmatrix{n \cr r}
\pmatrix{n \cr s}
\frac{ (-1)^t}{t+1} u^{r+t+1} v^s p^{-(r+s+t+1)}
\]
\noindent then
\[
\frac{1}{2\pi i} \oint \L^n \log\Big( 1+\frac{u}{p}\Big) \,\frac{dp}{p} =
\sum_{r+s+t=n-1}
\pmatrix{n \cr r}\pmatrix{n\cr s} \frac{ (-1)^t}{t+1} u^{n-s} v^s\,.
\]
\noindent The other parts of the integral may be evaluated in a
similar way and
so the final result is
\begin{eqnarray*}
{\widetilde Q}^{(n)} & = & + \frac{1}{2} \log(uv) \sum_{r+s=n}
\pmatrix{n \cr r}
\pmatrix{n \cr s} u^r v^s + 
\sum_{r+s+t=n-1}
\pmatrix{n \cr r}\pmatrix{n\cr s}
\frac{ (-1)^t}{t+1} ( u^{n-s} v^s +v^s u^{n-s} ) \\
& & - c_n
\sum_{r+s=n} \pmatrix{n \cr r} \pmatrix{n \cr s} u^r v^s\,.
\end{eqnarray*}

\bigskip

Note that there is an ambiguity in the definition of
${\widetilde Q}^{(n)}\,,$ since
${\widetilde Q}^{(n)} + \lambda Q^{(n)} $ is also a conserved charge.
The numbers
$c_n$ have the property that the coefficient of $(u^n+v^n)$ in the
terms not
involving $\log(uv)$ in the above expression is zero and this removes this
ambiguity. Explicitly this coefficient is
\[
\sum_{r+t=n-1} \pmatrix{n \cr r} \frac{ (-1)^t }{t+1} - c_n
\]
\noindent and this is, using the identity
\[
\sum_{t=0}^{n-m} \pmatrix{n \cr m+t} \frac{ (-1)^t}{t+1} =
\pmatrix{n \cr m-1}
\sum_{j=m}^n \frac{1}{j} \,,\quad\quad\quad n \geq m\,,
\]
\noindent zero, as required. Rearranging and resuming
(using the above identity) results in the
expression (\ref{eq:logcharges}).

\bigskip

\endproof

\bigskip

These logarithmic charges define a new set of evolution
equations defined by the
Lax equation

\[
\frac{\partial\L}{\partial\tau_n} =
\{ [ \L^n ( \log\L - c_n)]_{+}, \L \}\,,\quad
n=0\,,1\,,2\,,\ldots\,,
\]

\noindent or, equivalently, by the Hamiltonian equations

\[
u_{i,\tau_n} = \{ u_i , \widetilde{H}^{(n)} \}\,
\]

\noindent with $\widetilde{H}^{(n)}$  proportional to the logarithmic
charges $\widetilde{Q}^{(n)}\,.$ These flows all commute,
both with themselves and those defined
by (\ref{eq:lax}), and the corresponding Hamiltonians are in involution
with respect to the Hamiltonian structure defined by
(\ref{eq:hamstructure}).
The techniques developed in sections 2 and 4 may easily be extended to
derive these more complicated evolution equations explicitly. However, for
$N=2$ one may sidestep these calculations. It is straightforward to
show that
any flow that commutes with the basic equation (\ref{eq:mtoda}) must be of
the form

\begin{eqnarray*}
u_\tau & = & F u_x + u Q v_x \,, \\
v_\tau & = & F v_u + v Q u_x \,,
\end{eqnarray*}

\noindent where $F$ and $Q$ are any functions of $u$ and $v$ (not
necessarily symmetric) which satisfy the equations

\begin{equation}
\begin{array}{rcl}
F_u & = & v Q_v \,,\\
F_v & = & u Q_u \,,
\end{array}
\label{eq:logflux}
\end{equation}

\noindent or equivalently, the conservation law $Q_{t_1} = F_x\,.$ Thus
to find the evolution equations associated with the logarithmic charges
given by (\ref{eq:logcharges}) one only has to solve the above equations
for the flux $F\,.$

\bigskip

\noindent{\bf Lemma 7.2} The logarithmic charges (\ref{eq:logcharges})
give rise
to the evolution equations

\begin{eqnarray*}
u_{\tau_n} & = & \widetilde{F}^{(n)} u_x + u \widetilde{Q}^{(n)} v_x \,,\\
v_{\tau_n} & = & \widetilde{F}^{(n)} v_x + v \widetilde{Q}^{(n)} u_x \,,
\end{eqnarray*}

\noindent where the $\widetilde{Q}^{(n)}$ are the charges defined by
(\ref{eq:logcharges}) and the $\widetilde{F}^{(n)}$ are the corresponding
fluxes with respect to the $t_1$ flow, Explicitly

\begin{equation}
{\widetilde F}^{(n)} = \frac{1}{2} \log(uv) \sum_{r+s=n+1}
\pmatrix{n\cr r} \pmatrix{n\cr s} u^r v^s
-\sum_{r+s = n+1} \tilde{c}(r,s) u^r v^s
\label{eq:logflux2}
\end{equation}

\noindent where

\[
\tilde{c}(r,s) = \pmatrix{n\cr r} \pmatrix{n\cr s}
\Bigg[ c(r,s) + \frac{1}{1+n}\Bigg] -
\frac{1}{2(1+n)} \pmatrix{n+1\cr r} \pmatrix{n+1\cr s} \,.
\]

\bigskip

\noindent{\bf Proof}  Straightforward.
One just has to integrate equation (\ref{eq:logflux}).

\bigskip

\endproof

\bigskip

\noindent{\bf Example 7.3} The simplest such logarithmic flow is

\begin{eqnarray*}
u_{\tau_0} & = & \frac{1}{2} (u+v) u_x +
\frac{1}{2} u \log(uv) \, v_x \,,\\
v_{\tau_0} & = & \frac{1}{2} (u+v) v_x + \frac{1}{2} v \log(uv) \, u_x\,,
\end{eqnarray*}

\noindent or, in the original variables,

\begin{eqnarray*}
S_{\tau_0} & = & \frac{1}{2} S S_x + \frac{1}{2} \log P \, P_x \,, \\
P_{\tau_0} & = & \frac{1}{2} S P_x + \frac{P}{2} \log P \, S_x \,.
\end{eqnarray*}

\bigskip

\endexample

\bigskip

Calculations of this type were first performed (for $N=2$ and
small values of $n$)
in \cite{EguchiYang,EguchiYangHori} and then
(for $N=3$ and small values of $n$) in \cite{KannoOhta}.

\bigskip

\section*{8. Comments}

\bigskip

The results of this paper stem from the simple observation that by
factorizing the Lax function the calculations simplify dramatically,
As an example, this enables the metric which defines the
Hamiltonian structure to
be found for arbitrary $N\,.$ Even for $N=3$ the metric, when
written in terms of the
original, unmodified, variables is somewhat unmanageable and the
degree of complexity
increases rapidly as $N$ increases. Not all of the possible
formulae have been calculated
in this paper, though it should be clear how these may be derived.
For example, one
may regard the $\tau_n$ flows as negative $t_n$ flows
(i.e. $t_{-n}\equiv \tau_n$), and
so one may derive explicit forms for the function
$\Delta^{(m,n)}$ for all $m\,,n \in {\bf Z}\,.$
Similarly, the Hamiltonian structure for the evolution equations
derived from the
rational Lax function (\ref{eq:rationalLax}) has not been
calculated; to do this should
be straightforward.

\bigskip

The use of generalized hypergeometric functions to study the
conservation laws
follows from an earlier result \cite{FS} where it was shown that
the $N=2$ conservation
laws were expressible in terms of Legendre polynomials.
This results was used to construct
new families of integrable (or at least soluble) hierarchies
with similar properties to
the original Toda system, namely

\begin{eqnarray*}
u_t & = & u^{a-c} v^{c-1} \, v_x \,, \\
v_t & = & u^{a-c-1} v^{c} \, u_x \,,
\end{eqnarray*}

\noindent where $a$ and $c$ are free parameters.
The conservation laws of this system
depend on the hypergeometric function ${}_2 F_{1} (-n,n+a-1,c;z)\,.$
One interesting
possibility is to use the properties of the generalized
hypergeometric series to
perform a similar calculation for higher values of $N\,.$

\bigskip

As remarked at the very beginning of this paper, these systems come
from the dispersionless
limit of integrable dispersive systems. The Lax formalism for such
dispersive systems
comes from the replacements

\begin{eqnarray*}
p & \longrightarrow & \Delta \,, \\
\{f,g\} & \longrightarrow & [f,g]\,,
\end{eqnarray*}

\noindent where $\Delta$ is the shift operator introduced in section 1
and $[f,g]=fg-gf$
where now $f$ and $g$ are operators depending on $\Delta\,.$
Alternatively, one may use a
deformation of the Poisson bracket \cite{Strachan}, this having the
advantage that one retains
a geometrical description of these hierarchies \cite{TakasakiTakebe}.
While all this
clearly works for polynomial Lax functions, there is a problem of
how to find dispersionless
analogues of the systems derived from the rational Lax function
(\ref{eq:rationalLax}), the
difficulty coming form the interpretation of the operator
$(\Delta+u)^{-1}\,.$ One may to
resolve this problem has been recently proposed in \cite{Orlov}.

\bigskip

Finally, it must be stressed again that the {\sl existence} of the
results in this paper have
been known for many years. However, apart from some trivial examples,
the precise forms of
these results have not been calculated. We have adopted an utilitarian
view that
such implicit formulae should be calculated explicitly, and the use of
modified variables,
as advocated in this paper, provides a simple way to do this.

\bigskip

\bigskip

\section*{Acknowledgements}

I.A.B.S. would like to thank the University of Newcastle, where part of
this paper was written,
for the Wilfred Hall fellowship.


\bigskip

\bigskip





\begin{thebibliography}{**}
%
\bibitem{Kuperschmidt}{B.A. Kuperschmidt, {\sl Ast\'erisque}, {\bf 123}
(1985) 1-212.}
%
\bibitem{TakasakiTakebe}{K. Takasaki and T. Takebe,
{\sl Integrable hierarchies
and the dispersionless limit}, preprint UTMA 94-35, hep-th/9405096.}
%
\bibitem{Dubrovin}{B.A. Dubrovin, {\sl Geometry of 2D topological
field theories},
preprint, hep-th/9407018.}
%
\bibitem{FS}{D.B. Fairlie and I.A.B. Strachan, {\sl Physica D} {\bf 90}
(1996) 1-8.}
%
\bibitem{AoyamaKodama}{S. Aoyama and Y. Kodama, {\sl Topological
Landau-Ginzburg
theory with a rational potential and the dispersionless KP hierarchy},
preprint hep-th/9505122.}
%
\bibitem{benney}{D.J. Benney, {\sl Stud.Appl. Math.} {\bf 52} (1973),
45-50.}
%
\bibitem{pc}{G. Watts, private communication.}
%
\bibitem{CarrollKodama}{R. Carroll and Y. Kodama, {\sl J. Phys. A}
{\bf 28} (1995)
6373-6387.}
%
\bibitem{Son}{S.H. Son, {\sl The equations of some dispersionless limit},
preprint hep-th/9506172.}
%
\bibitem{Gelfand}{I.M Gel'fand, M.I. Graev and V.S. Retakh,
{\sl Uspekhi Mat. Nauk} {\bf 47}
(1992) 3-82.}
%
\bibitem{DubrovinNovikov}{B.A. Dubrovin and S.P. Novikov,
{\sl Sov. Math. Dokl.}
{\bf 27} (1983) 665.}
%
\bibitem{Tsarev}{S.P. Tsarev, {\sl Sov. Math. Dokl.} {\bf 31} (1985) 488;
{\sl Math. USSR Izves.} {\bf 37} (1991) 397.}
%
\bibitem{Bureau}{F.J. Bureau, {\sl Annali di Mathematica pura ed
applicata}, {\bf 94}
(1972) 345-360; {\sl Bull.de l'Acad\'emie royale de Belgique}, {\bf 73}
(1987) 335-353.}
%
\bibitem{GibbonsTsarev}{J. Gibbons and S.P. Tsarev, {\sl Reductions of
the Benney
Equations}, preprint, Nov. 1995.}
%
\bibitem{EguchiYang}{T.Eguchi and S. Yang, {\sl Mod. Phys. Lett. A}
{\bf 9} (1994) 2893-2902.}
%
\bibitem{EguchiYangHori}{T. Eguchi, K. Hori and S. Yang,
{\sl Topological $\sigma$-models
and large-$N$ Matrix Integrals.}, preprint hep-th/9503017.}
%
\bibitem{KannoOhta}{H. Kanno and Y. Ohta, {\sl Nucl. Phys. B} {\bf 442}
(1995) 179-204.}
%
\bibitem{Strachan}{I.A.B. Strachan, {\sl A Geometry for Multidimensional
Integrable Systems},
preprint hep-th/9604142, to appear {\sl J. Geom. Phys..}}
%
\bibitem{Orlov}{B. Enriquez, A. Yu. Orlov and V.N. Rubtsov,
{\sl Inverse Problems,} {\bf 12}
(1996) 241-250.}
%
\end{thebibliography}
\end{document}